# Interactions of space-variant polarization beams with Zeeman-shifted rubidium vapor

Anat Szapiro,[1] Liron stern,[1] and Uriel Levy[1,*]

[1]*Department of Applied Physics, The Benin School of Engineering and Computer Science, The Hebrew University of Jerusalem, Jerusalem, 91904, Israel*
*\*Corresponding author: ulevy@cc.huji.ac.il*

Space variant beams are of great importance as a variety of applications have emerged in recent years. As such, manipulation of their degrees of freedom is highly desired. Here, by exploiting the circular dichroism and circular birefringence in a Zeeman-shifted Rb medium, we study the general interaction of space variant beams with such a medium. We present two particular cases of radial polarization and hybrid polarization beams where the control of the polarization states is demonstrated experimentally. Moreover, we show that a Zeeman-shifted atomic system can be used as an analyzer for such space variant beams. © 2014 Optical Society of America
OCIS Codes: (140.3295) Laser beam characterization (260.5430) Polarization (160.3820) Magneto-optical materials (260.7490) Zeeman Effect (300.6210) Spectroscopy, atomic

In recent years, the optical properties of space-variant polarized beams, that is, beams that have several spatial inhomogeneous polarization states, have drawn significant attention and interest and have been the topic of numerous theoretical and experimental investigations in recent years due to the variety of their potential applications, including tight focus and optical trapping [1,2], particle orientation analysis [3], single shot polarization-dependent measurements [4], etc. Recently, the interaction of space variant polarized beam and space variant polarized atoms has been explored [5, 6]. Atomic vapor cells consisting of Rubidium (Rb) or Cesium (Cs) are used extensively both in industry and academia for a myriad of applications [7,8]. Rb has a number of optical transitions at convenient wavelengths. This, together with the capability of obtaining high optical densities in temperatures close to room temperature and its high Verdet constant [9], such that the Rb atoms exhibit a strong response to an externally applied magnetic field, makes Rb a fine candidate for exploring and manipulating optical fields. Here, we use the Zeeman-shifted Rb as a frequency-dependent dichroic and birefringent medium (see schematic in Fig. 1). Specifically, we use a Rb cell to manipulate the spatial distribution and polarization properties of the vector beams. Furthermore, the Rb cell is used as an efficient probe that enables direct imaging of the local state of polarization across the beam.

When atomic systems such as Rb vapor are subjected to the influence of an external magnetic field, the degenerated energy levels of the hyperfine structure of the atoms, defined by the quantum number F, split into 2F+1 magnetic sub-levels. The splitting of these energy levels changes their absorption spectra. This, together with the predefined selection rules, makes the Rb medium highly dichroic and birefringent for right-hand circular polarization (RHC) and left-hand circular polarization (LHC).

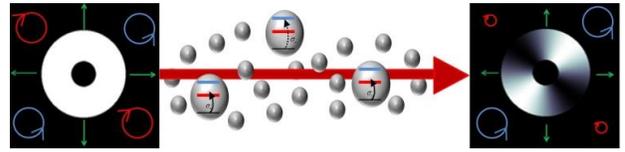

Fig.1: A schematic figure showing space-variant polarized beams interacting with Zeeman-shifted Rb atoms.

Fig.2 shows the numerical calculation of (a) the absorption coefficient, and (b) the refractive index of $^{85}$Rb D2 F=2 to F'=1, 2, 3 Doppler broadening transitions under the influence of a magnetic field (B=200 Gauss). As discussed previously, the magnetic field creates different frequency dependences for both the absorption coefficient and refractive index for RHP (red) and LHP (black). Fig. 2.a(i) shows the divergence of the absorption coefficient between RHP and LHP as a function of frequency, where this divergence creates circular dichroism (CD) frequency-dependent media. Fig. 2.b(ii) shows the divergence of the refractive index between RHP and LHP as a function of frequency, where this divergence creates a circularly birefringent (CB) frequency-dependent medium. In the calculation below, we used a linear Zeeman regime approximation where the shift in the energy levels can be written as $\Delta E \cong -\mu_B g_F B$, where $\mu_B$ is the Bohr magneton, $g_F$ is the Lande g-factor and $B$ is the applied magnetic field. We note that this approximation is relevant as long as the Zeeman shift split is smaller than the hyperfine energy structure. In the experimental work presented, the excited states ($5^2P_{1/2}$) exceed the linear approximation regime for several of the applied magnetic field values. Nevertheless, the physical mechanism remains the same. The zero frequency detuning was set to be the resonant point of the zero magnetic field.

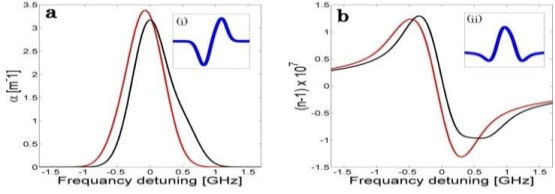

Fig.2: A numerical calculation of the frequency dependent a) absorption coefficient and b) refractive index of Zeeman-shifted $^{85}$Rb D2 F=2 to F'=1, 2, 3 Doppler broadening transitions for RHC (red) and LHC (black) in a magnetic field of B=200G. The inner graphs, (i) and (ii), show the difference in extinction coefficient and refractive index (respectively) between RHP and LHP.

In the absence of a magnetic field, the RHP and LHP extinction coefficient and refractive index converge. As evident in Fig.2, dichroism and birefringence are enhanced by the application of a magnetic field. These two phenomena have strong yet different frequency dependences. For example, in the small region of zero frequency detuning almost no dichroism is seen (zero change in the absorption coefficient). Nevertheless, in the same frequency detuning region a maximum divergence of the refractive index between RHP and LHP occurs and this creates significant CB in the medium. A small change in the detuned frequency (~0.25 GHz) will drastically change this picture by creating a different ratio between the CD and CB, and this strong frequency-dependence allows us to manipulate the amount of CB and CD in the medium. Next, we describe the interactions of space-variant polarization with such CD and CB media. A general field of space-variant angular dependent polarized light can be written as $E(\theta) = \widetilde{g(\theta)}\hat{\sigma}_+ + \widetilde{f(\theta)}\hat{\sigma}_-$, where $g$ and $f$ are general complex functions of the angular polar coordinate $\theta$, $\hat{\sigma}_+$ and $\hat{\sigma}_-$ are the Jones vectors of RHP and LHP. In this work we focus on two types of space-variant polarization beams: The first is the radially polarized beam, which is a beam of light with linear polarizations directed towards the center of the beam. In general, the electric field of a radially polarized beam undergoing CB and CD can be described by the following equation:

$$E(\theta,\phi,a) \sim [ae^{i(\theta+\phi)}\hat{\sigma}_+ + e^{-i\theta}\hat{\sigma}_-] \quad (1)$$

where $a$ represents the CD – the relative difference in absorption between RHC and LHC, $\phi$ represents the induced phase difference between these two polarizations, which causes CB to occur. The second type of beam investigated in this work is the hybrid polarization beam (HPB). These beams are composed of spatially separated linear, elliptical, right-hand and left-hand circular polarization. HPB can be represented in the general form of

$$E(\delta,\phi,a) \sim [ae^{i\phi}(e^{i\delta} - ie^{-i\delta})\hat{\sigma}_+ + (e^{i\delta} + ie^{-i\delta})\hat{\sigma}_-] \quad (2)$$

where $a$ is the CD, $\phi$ represents the phase difference between RHC and LHC, $\delta = m\theta + \frac{2\pi nr^2}{r_0} + \theta_0$, where m is a nonnegative integer that represents the azimuthal index, the arbitrary number n is the radial index, $r_0$ is the radius of the vector field, $\theta_0$ is the initial phase and $r, \theta$ are the spatial coordinates in a cylindrical coordinate system [10]. Using equations (1) and (2) we plot three different polarization maps with and without the influence of CB and CD (Fig.3).

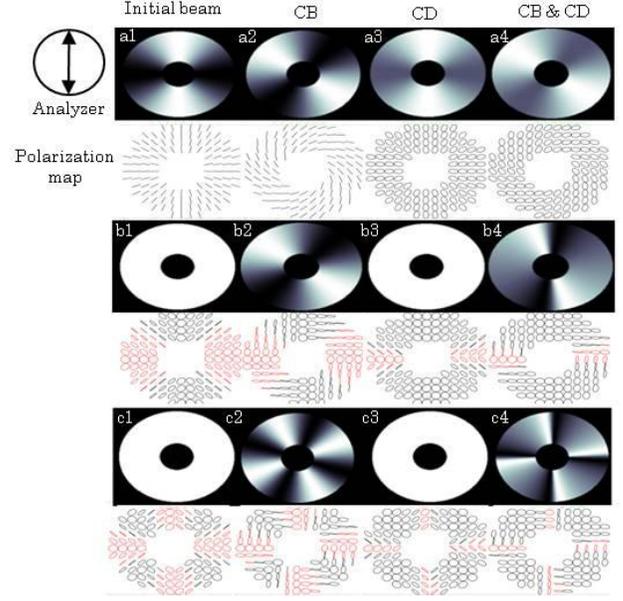

Fig.3: Numerical calculations of the intensity and polarization states for three different space-variant beams, corresponding to CB, CD and combined media. The letter index represents the type of space-variant vector beam: a) radial b) hybrid with m=1, n=0 and c) hybrid with m=2, n=0. The top panel of each type of beam represents the intensity distribution and bottom panel represents the polarization map. The colors in the polarization maps represent the polarization orientation- red and black for RHP and LHP respectively

Fig. 3 predicts the influence of a Zeeman-shifted Rb medium on space-variant beams. As can be seen in column 2 of Fig. 3, the CB medium rotates the polarization direction homogenously in space while the beam maintains its polarization distribution – i.e. the linear, circular or elliptical contents remain. This is the opposite of the CD medium, which maintains the polarization direction of the transmitted beam while changing its polarization characteristics. For example, Fig. 3.a2 demonstrates the homogenous linear polarization rotation of the beam caused by the CB medium, whereas Fig 3.a3 shows the polarization transition from linear to elliptical produced by the CD medium. In this case, the orientation of both principal axes of the elliptical polarization remains unchanged with respect to the original beam. Clearly, the influence of both CD and CB media is a combination of changes in the polarization character and rotation of the polarization direction as can be seen in Fig. 3 column 4. We note that in panels b1, b3, c1 and c3 a constant intensity distribution is observed due to the symmetry of the beam with respect to the analyzer direction.

Fig. 4 shows a scheme of the experimental set-up. A Tunable Diode Laser (DL 100, Toptica) was scanned around the Rb D1 transition. We used a radial polarizer converter (RPC, Altechna) to create radially polarized beams; the generation of a hybrid polarization beam with *m=1* and *n=0* was established by propagating the radially polarized beam through a quarter wave plate (QWP) [11]. To find the space variant polarization states of the beam, the Stokes parameters measurement method was carried out by taking four independent measurements using a QWP, a linear polarizer (analyzer) and a CCD detector. An external magnetic field was induced by applying current through a home-made copper solenoid to create a magnetic field $B \approx 180G$ (measured by a Gauss meter). The Rb cell was heated to $T \approx 70^0 C$ in order to enhance the atom-light interaction.

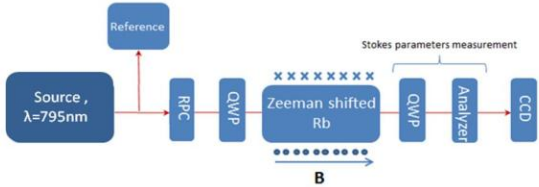

Fig.4: Sketch of the experimental setup. A radial polarization converter (RPC) and a quarter wave plate (QWP) situated in front of the Rb cell generate radially- and hybrid polarized beams. The QWP and the analyzer beyond the Rb cell are used for Stokes parameters measurement and analysis.

The Faraday Effect is a specific example of linearly-polarized beam interaction with a CB medium. Whenever a linearly polarized beam enters a CB medium, the difference in optical path length between RHP and LHP causes polarization rotation as a function of the applied magnetic field. Naturally, the same physical mechanism applies for radially polarized beams as mentioned previously and shown in Fig. 3.a2. This example is worth special attention as the growing interest in radially polarized beams and their applications raises the need for new generations of space-variant customized optical tools, one of which is a radial optical isolator. The Faraday effect is the main physical mechanism required to create an optical isolator, and implementation of the Faraday effect on radially polarized beams is the first step towards creating a radial optical isolator. Fig. 5 shows the measured radial Faraday Effect. As seen in the figure, Faraday effect measurements were taken in the regime where the dichroism is minor and CB is the dominant optical phenomenon. A nearly 45 degree polarization rotation is observed. The radial Faraday effect measurements are also a convenient method for calculating the angle of the polarization rotation in a single measurement. The rotation angle of the polarization converges with the rotation angle of the symmetry axis of the beam intensity (Fig. 5, top). This is a unique property of the space variant polarized beam.

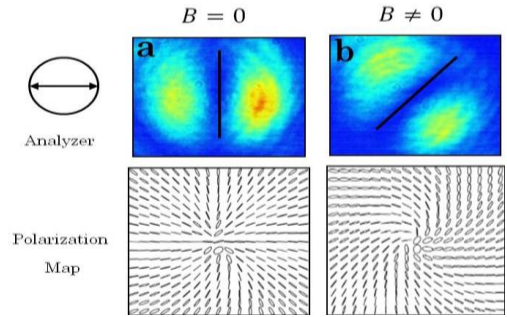

Fig.5: Measurements of radially polarized beam intensity with analyzer (top) and polarization map (bottom) with (b) and without (a) the influence of Zeeman-shifted Rb.

Fig. 6 shows the hybrid polarized beam (*m=1*, *n=0*) interacting with Zeeman-shifted Rb in the near resonance D1 transition regime. Different wavelength regimes show different ratios between CD and CB in the media, as described below.

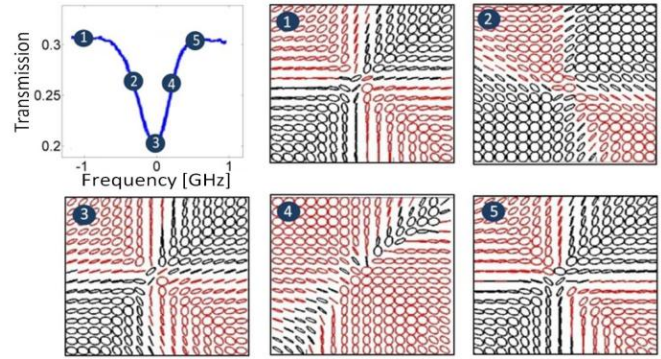

Fig.6: Stokes parameters measurements of a hybrid polarized beam transmitted through the Zeeman-shifted Rb medium at different wavelengths. Red is for RHP and black is for LHP. The top left panel shows the absorption spectrum of the reference cell without external magnetic field and the relevant measured points.

Fig. 6(1) and Fig. 6(5) show practically no influence of either CD or CB on the transmitted beam, as expected in the off-resonance regime. Indeed, the initial beam polarization map of the transmitted beam before entering the Zeeman-shifted Rb medium (not shown) resembles almost exactly the maps represented in Fig. 6 (1) and Fig. 6(5). In Fig. 6(2) we see a clear CD effect; in this regime the RHP has a higher absorption coefficient and thus the major polarizations of the transmitted beam are LHP. Fig. 6(4) presents a similar picture yet with the opposite dichroism. In Fig. 6(3), CB is noticeable (about $30^0$ of polarization rotation) compared to the off-resonance polarization maps and CD is barely observed. Note that in these figures the magnitude of the intensity is not presented, yet naturally, the intensity of the beam shown in Fig. 6(3) is smaller than in the other figures, as it is located completely within the resonance regime and thus a higher percentage of the beam's intensity is absorbed.

Finally, we focus on the intensity distribution of the beam as a function of frequency. Fig. 7 shows the spatial intensity measurements of HPB transmitted through the Zeeman-shifted Rb medium. Note that these measurements were taken without the analyzer and QWP.

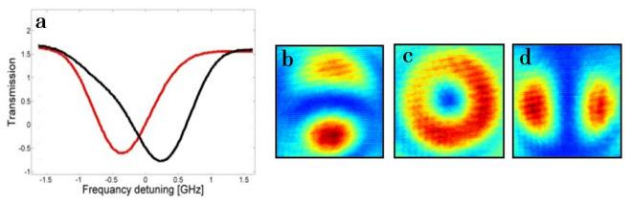

Fig.7: (a) measured absorption spectrum of a Zeeman-shifted Rb medium in the $^{85}$Rb transition manifold for RHP and LHP (red and black, respectively). (b-d) measured transmission intensity of the HPB at various wavelengths; (b) in the regime of RHP resonance; (c) in the regime of equivalent absorption between RHP and LHP; (d) regime of LHP resonance.

At the frequency region of RHP resonance (Fig. 7b), the RHP regions in the beam are absorbed. Similarly, when the frequency of the beam is near the LHP resonance (Fig. 7d) the LHP regions in the beam are absorbed and thus the intensity map is shown to be rotated at 90 with respect to the previous case. Finally, when the absorption coefficients of RHP and LHP are equal (Fig. 7c), the absorption in the beam is homogeneous, and the intensity ratio of the beam is maintained. The Zeeman-shifted Rb in these measurements is used as an analyzer of the HPB, and by detecting the beam intensity at different frequencies we can learn about the spatial polarization distribution of the beam.

In summary, we calculated the state of polarization for space- variant polarization beams for three cases: Radially polarized beam, HPB with m=1; and HPB with *m=2*. We have calculated the spatial polarization distribution of these beams after being transmitted through CB and CD media. We measured the spatial polarization distribution of a radial beam and HPB with *m=1* passing through a Zeeman-shifted Rb medium. Next, we showed the use of the Rb medium as an analyzer for space-variant polarization beams.

These demonstrated results indicate that the interaction of hot vapors with space variant polarized beams provides additional degrees of freedom in controlling and manipulating the properties of both the beam and the atomic media. Therefore, the demonstrated approach may be used in applications such as isolators for radially polarized light, and probing the uniformity of magnetic fields, to name a few.